\documentclass[sigconf]{acmart}

\AtBeginDocument{%
  \providecommand\BibTeX{{%
    \normalfont B\kern-0.5em{\scshape i\kern-0.25em b}\kern-0.8em\TeX}}}

\setcopyright{acmcopyright}
\copyrightyear{2018}
\acmYear{2018}
\acmDOI{10.1145/1122445.1122456}

\acmJournal{TOG}
\acmVolume{37}
\acmNumber{4}
\acmArticle{111}
\acmMonth{8}

\usepackage{listings}
\usepackage{xcolor}
\usepackage{graphicx}
\usepackage[export]{adjustbox}
\usepackage{xspace}
\usepackage{tikz}
\usepackage{multirow}
\usepackage{array}
\usepackage{algorithm}
\usepackage{algpseudocode}
\definecolor{codegreen}{rgb}{0,0.6,0}
\definecolor{codegray}{rgb}{0.5,0.5,0.5}
\definecolor{codepurple}{rgb}{0.58,0,0.82}
\definecolor{backcolour}{rgb}{0.95,0.95,0.92}

\lstdefinestyle{mystyle}{
    backgroundcolor=\color{backcolour},   
    commentstyle=\color{codegreen},
    keywordstyle=\color{magenta},
    numberstyle=\tiny\color{codegray},
    stringstyle=\color{codepurple},
    basicstyle=\ttfamily\footnotesize,
    breakatwhitespace=false,         
    breaklines=true,                 
    captionpos=b,                    
    keepspaces=true,                 
    numbers=left,                    
    numbersep=5pt,                  
    showspaces=false,                
    showstringspaces=false,
    showtabs=false,                  
    tabsize=2
}

\renewcommand\footnotetextcopyrightpermission[1]{} 

\newcolumntype{P}[1]{>{\centering\arraybackslash}p{#1}}
\acmConference[]{}{}{}

\begin{document}

\title{\mytool: A Dataset of Real World Ethereum Smart Contracts Labelled with Vulnerabilities}

\author{Chavhan Sujeet Yashavant}
\affiliation{%
  \institution{Indian Institute of Technology Kanpur}
  \city{Kanpur}
  \country{India}}
\email{sujeetc@cse.iitk.ac.in}

\author{Saurabh Kumar}
\affiliation{%
  \institution{Indian Institute of Technology Kanpur}
  \city{Kanpur}
  \country{India}}
\email{skmtr@cse.iitk.ac.in}

\author{Amey Karkare}
\affiliation{%
  \institution{Indian Institute of Technology Kanpur}
  \city{Kanpur}
  \country{India}}
\email{karkare@cse.iitk.ac.in}


\definecolor{verylightgray}{rgb}{.97,.97,.97}

\lstdefinelanguage{Solidity}{
	keywords=[1]{address, anonymous, assembly, assert, balance, break, call, callcode, case, catch, class, constant, continue, constructor, contract, debugger, default, delegatecall, delete, do, else, emit, event, experimental, export, external, false, finally, for, function, gas, if, implements, import, in, indexed, instanceof, interface, internal, is, length, library, log0, log1, log2, log3, log4, memory, modifier, msg, new, payable, pragma, private, protected, public, pure, push, require, return, returns, revert, selfdestruct, send, sender, solidity, storage, struct, suicide, super, switch, then, this, throw, transfer, true, try, typeof, using, value, view, while, with, addmod, ecrecover, keccak256, mulmod, ripemd160, sha256, sha3}, 
	keywordstyle=[1]\color{blue}\bfseries,
	keywords=[2]{bool, byte, bytes, bytes1, bytes2, bytes3, bytes4, bytes5, bytes6, bytes7, bytes8, bytes9, bytes10, bytes11, bytes12, bytes13, bytes14, bytes15, bytes16, bytes17, bytes18, bytes19, bytes20, bytes21, bytes22, bytes23, bytes24, bytes25, bytes26, bytes27, bytes28, bytes29, bytes30, bytes31, bytes32, enum, int, int8, int16, int24, int32, int40, int48, int56, int64, int72, int80, int88, int96, int104, int112, int120, int128, int136, int144, int152, int160, int168, int176, int184, int192, int200, int208, int216, int224, int232, int240, int248, int256, mapping, string, uint, uint8, uint16, uint24, uint32, uint40, uint48, uint56, uint64, uint72, uint80, uint88, uint96, uint104, uint112, uint120, uint128, uint136, uint144, uint152, uint160, uint168, uint176, uint184, uint192, uint200, uint208, uint216, uint224, uint232, uint240, uint248, uint256, var, void, ether, finney, szabo, wei, days, hours, minutes, seconds, weeks, years},	
	keywordstyle=[2]\color{teal}\bfseries,
	keywords=[3]{block, blockhash, coinbase, difficulty, gaslimit, number, timestamp, , data, gas, sig, value, now, tx, gasprice, origin},	
	keywordstyle=[3]\color{violet}\bfseries,
	identifierstyle=\color{black},
	sensitive=false,
	comment=[l]{//},
	morecomment=[s]{/*}{*/},
	commentstyle=\color{gray}\ttfamily,
	stringstyle=\color{red}\ttfamily,
	morestring=[b]',
	morestring=[b]"
}

\lstset{
	language=Solidity,
	backgroundcolor=\color{verylightgray},
	extendedchars=true,
	basicstyle=\ttfamily, 
	showstringspaces=false,
	showspaces=false,
	numbers=left,
	numberstyle=\footnotesize,
	numbersep=9pt,
	tabsize=2,
	breaklines=true,
	showtabs=false,
	captionpos=b
}


\newcommand{\mytool}{{\scshape ScrawlD\xspace}}
\newcommand{\numtools}{5\xspace}

\newcommand{\totalSCInOurDB}{\textbf{6.7K} } 
\newcommand{\numberTotalSCInOurDB}{\textbf{6780} }

\newcommand{\crossmark}{\ensuremath{\times}}
\newcommand{\unknown}{--}


\begin{abstract}

Smart contracts on Ethereum handle millions of U.S. Dollars and other financial assets. In the past, attackers have exploited smart contracts to steal these assets.  The Ethereum community has developed plenty of tools to detect vulnerable smart contracts. However, there is no standardized data set to evaluate these existing tools, or any new tools developed. There is a need for an unbiased standard benchmark of real-world Ethereum smart contracts. We have created \mytool: an annotated data set of real-world smart contracts taken from the Ethereum network. The data set is labelled using \numtools~tools that detect various vulnerabilities in smart contracts, using majority voting. 

\end{abstract}



\maketitle

\section{Introduction}
Permissionless blockchains like Ethereum have tens of millions of smart contracts~\cite{Smart-Contracts-and-Opportunities-for-Formal-Methods}. A smart contract is a code that resides on the Ethereum network and executes when predetermined conditions get satisfied. Smart contracts on Ethereum handle financial assets worth millions of U.S. Dollars (USD). Attacking smart contracts can provide monetary benefits to the attackers or can cause other damages~\cite{A-Survey-of-Attacks-on-Ethereum-Smart-Contracts}. Therefore, the attacks on Ethereum smart contracts have increased rapidly over the last few years~\cite{A-Survey-on-Ethereum-Systems-Security-Vulnerabilities-Attacks-and-Defenses}. 

To detect the attacks on Ethereum smart contracts, the Ethereum community has developed plenty of tools~\cite{slither,oyente,securify,smartcheck,mythril,maian}. These tools analyze smart contracts and produce vulnerability reports. The authors of these tools use datasets of smart contracts to evaluate the tools' efficiency, correctness, and other parameters. However, datasets used by the authors to evaluate their tools vary significantly~\cite{Empirical-Evaluation-of-Smart-Contract-Testing:What-is-the-Best-Choice}. The type of dataset has an impact on the tool's performance. Therefore it is necessary to evaluate tools using multi-type integrated benchmark suite~\cite{Empirical-Evaluation-of-Smart-Contract-Testing:What-is-the-Best-Choice}. 

Ren et al.~\cite{Empirical-Evaluation-of-Smart-Contract-Testing:What-is-the-Best-Choice} have created a dataset having has two types of smart contracts: annotated and non-annotated. However, the number of annotated smart contracts in the database is very small (less than a thousand). Further, most of these annotated smart contracts not real-world contracts. The dataset created by Ferreira et al.~\cite{smartbugs} the same issues. Hence, these datasets cannot be considered representatives for the real-world Ethereum smart contracts. There is a need for a benchmark suite containing real world smart contracts labelled with vulnerabilities. 

We present \mytool: a dataset of real-world Ethereum smart contracts labelled with vulnerabilities. The Ethereum community can use \mytool~for an unbiased evaluation of new and existing tools to analyse vulnerabilities.\mytool~contains \totalSCInOurDB labelled real-world Ethereum smart contracts. Our methodology to label the smart contracts is based on the work by Ren~et.~al.~\cite{Making-Smart-Contract-Development-More-Secure-and-Easier}. This paper describes our solution that uses various tools (such as~\cite{slither,smartcheck,mythril,oyente,osiris}) to find and catalog the vulnerabilities in the smart contracts. The resulting data set, \mytool, is released at \url{https://github.com/sujeetc/ScrawlD}.

\section{Related Work}
\label{sec:relwork}

Ren et al.~\cite{Empirical-Evaluation-of-Smart-Contract-Testing:What-is-the-Best-Choice} crafted a benchmark suite that integrates labeled and unlabeled Smart Contracts. The unlabeled dataset has 45,622 Real-World Ethereum Smart Contracts having more than one transaction on the Ethereum blockchain. The labeled dataset has artificially constructed contracts (350 contracts) and confirmed vulnerable contracts (214 contracts). 

Durieux et al.~\cite{Empirical-Review-of-Automated-Analysis-Tools-on-47587-Ethereum-Smart-Contracts} crafted a dataset of annotated and non-annotated smart contracts. The annotated part contains 69 contracts tagged with 115 vulnerabilities. The annotated part has ten categories of vulnerabilities. In contrast, the non annotated part contains 47,518 unique contracts, each with at least one transaction on the Ethereum network. 

Researchers can't use the unlabeled dataset to evaluate the tool as there is no ground truth associated with it. Moreover, the labeled dataset doesn't have more than a thousand smart contracts.   

Many vulnerability analysis tools used unlabelled datasets to evaluate the tool. Due to huge number of contracts in these datasets, either the dataset is not annotated or a very small subset is annotated with the vulnerabilities present. To the best of our knowledge, there is no fully annotated dataset released publicly. We describe existing efforts in this area. 
 Smartcheck~\cite{smartcheck} authors evaluate the tool's accuracy by using three annotated smart contracts. These annotated smart contracts do not represent the diverse smart contracts from Ethereum. Moreover, the authors used 4,600 verified contracts from Ethereum to evaluate the tool's efficiency. However, the issues found in these 4,600 smart contracts are not manually verified.

 Slither~\cite{slither} does two types of experiments to evaluate the tool. The first experiment contains DAO~\cite{dao} and SpankChain~\cite{spankchain} smart contracts. DAO and SpankChain are two famous reentrancy attacks on Ethereum. The second experiment contains 1000 most used smart contracts from the Ethereum network. This dataset is unlabelled. The authors manually reviewed the results of detecting  reentrancy vulnerability  for 50 random smart contracts to evaluate the accuracy of Slither. Thus, their validation is not representative of the possible vulnerabilities in a dataset having large number of smart contracts.

\section{Methodology}

\begin{table*}[t]
  \caption{Vulnerabilities Description and SWC-ID as per SWC Registry
  }
  \label{table:vuln_description}
 \renewcommand{\arraystretch}{1.2}
  \begin{tabular}{ccc}
    \toprule
            
    \textbf{Vulnerability Type} & \textbf{SWC-ID} & \textbf{Description} \\
    \midrule
    ARTHM & SWC-101 & Arithmetic (Integer Overflow and Underflow) \\
    DOS & SWC-113, SWC-128 & Denial of Service \\
    LE & -- & Locked Ether \\
    RENT & SWC-107 & Reentrancy \\
    TimeM & SWC-116 & Block values as a proxy for time \\
    TimeO & SWC-114 & Transaction Order Dependence \\
    UE & SWC-104 & Unchecked Call Return Value \\
    TX-Origin & SWC-115 & Authorization through tx.origin \\
\bottomrule
\end{tabular}
\end{table*}

\begin{table*}[t]
  \caption{Vulnerabilities Supported by Selected Tools.}
  \label{tab:freq}
 \renewcommand{\arraystretch}{1.2}
  \begin{tabular}{ccccccccc}
  
  \multicolumn{9}{l}{\checkmark~indicates that the tool supports the vulnerability.}\\
  \multicolumn{9}{l}{\crossmark~indicates that the tool doesn't support the vulnerability.}\\
  \multicolumn{9}{l}{\_~indicates Unknown.} \\ \toprule

\multirow{2}{*}{\textbf{Tool Name}} & \multicolumn{8}{c}{\textbf{Vulnerability Type}} \\
   & ARTHM & DOS & LE & RENT & TimeM & TimeO & TX-Origin & UE \\                 \midrule
    Slither    & \crossmark  &  \checkmark  & \checkmark  &  \checkmark  & \checkmark  & \crossmark  & \checkmark  & \checkmark \\
    Smartcheck & \checkmark  & \checkmark  & \checkmark  & \crossmark   & \checkmark & \crossmark  & \checkmark & \checkmark \\
    Mythril    & \checkmark  & \checkmark  &  \unknown  & \checkmark  & \checkmark  & \checkmark  & \checkmark  & \checkmark \\
    Oyente    & \checkmark & \unknown & \crossmark  &  \checkmark & \checkmark  & \checkmark  &   \unknown & \crossmark \\
    Osiris  & \checkmark & \crossmark  & \crossmark & \crossmark & \crossmark & \crossmark & \crossmark & \crossmark  \\
    \bottomrule
\end{tabular}
\label{table:vuln_supported_per_tool}
\end{table*}

\begin{table}[t]
  \caption{Threshold for each Vulnerability Type}
  \label{table:threshold}
  \begin{tabular}{P{0.2\linewidth}P{0.4\linewidth}P{0.2\linewidth}}
    \toprule
    \textbf{Vulnerability Type} & \textbf{\#Tools that can detect the Vulnerability} & \textbf{Threshold} \\
    \midrule
    ARTHM & 4 & 2 \\
    DOS & 3 & 2 \\
    LE & 2 & 1 \\
    RENT & 3 & 2 \\
    TimeM & 4 & 2 \\
    TimeO & 2 & 1 \\
    UE & 3 & 2 \\
    TX-Origin & 3 & 2 \\
\bottomrule

\end{tabular}

\end{table}

This section explains the data source, tools used, and process to label the dataset with an example. 

\subsection{Data Source}
We took 45,622 addresses belonging to smart contracts from the dataset of the paper by Ren~et.~al.~\cite{Empirical-Evaluation-of-Smart-Contract-Testing:What-is-the-Best-Choice} The dataset contains all unique smart contracts with more than one transaction on the Ethereum network. Moreover, these smart contracts' source code is publicly available. We crawled the source codes of these 45,622 smart contracts through Etherscan's API~\cite{etherscan}. However, we released~\numberTotalSCInOurDB smart contracts in \mytool~due to challenges given in Section~\ref{section:challenges}.

\subsection{Tools Used}

The following points explain the criteria to select the tools as used in the paper by Durieux ~et.~al.~\cite{Empirical-Review-of-Automated-Analysis-Tools-on-47587-Ethereum-Smart-Contracts}

    \begin{itemize}
        \item \textbf{Available and CLI:} The tool is released publicly and has a command-line interface (CLI).
        \item \textbf{Compatible Input:} The tool requires input as Solidity smart contract. We do not consider the tools that require only EVM bytecode.
        \item \textbf{Only Source:} The tool runs on only Solidity source code. We do not consider the tools that require a test suite or smart contract labelled with assertions. 
        \item \textbf{Vulnerability Finding:} The tool finds vulnerabilities in smart contracts. 
    \end{itemize}

The following tools are selected based on the above criteria.
\begin{itemize}
    \item \textbf{Slither:} Slither~\cite{slither} is a static analysis tool designed to analyze Ethereum smart contracts. It has four prominent use cases: automated detection of vulnerabilities, automated detection of code optimization opportunities, improvement of users' understanding of the contracts, and assistant with code review.
    \item \textbf{Mythril:} Mythril is a tool that does security analysis of Ethereum smart contracts. It detects various security issues~\cite{mythril}.
    \item \textbf{Smartcheck:} Smartcheck~\cite{smartcheck} is an extensible static analysis tool that detects vulnerabilities in smart contracts. It converts Solidity code into XML-based intermediate representation and checks it with XPATH patterns. 
    
    \item \textbf{Oyente:} Oyente~\cite{oyente} is a symbolic execution tool to find security bugs in smart contracts. 
    
    \item \textbf{Osiris:} Osiris~\cite{osiris} is a framework comprising symbolic execution and taint analysis. It detects integer bugs in Ethereum smart contracts. 
\end{itemize}

These tools are neither sound nor complete. They can miss vulnerabilities (False Negatives) or detect them when they are not present (False Positives).

\subsection{Vulnerabilities}

Table \ref{table:vuln_description} shows vulnerabilities selected for evaluation. The column ``Vulnerability type'' indicates the acronym used in the paper for a particular vulnerability, while the column ``SWC-ID'' shows the ID provided by the SWC Registry~\footnote{SWC Registry: \url{https://swcregistry.io}} for the vulnerability where available. 

Table \ref{table:vuln_supported_per_tool} shows vulnerabilities supported by selected tools. Symbol \checkmark indicates that the tool can detect a particular vulnerability, whereas symbol \crossmark indicates that the tool can't detect the vulnerability. Symbol \unknown~indicates that we don't know whether the tool has detection capability of the particular vulnerability or not. Due to improper documentation of the tool, it is hard to find whether it supports this vulnerability or not.

\subsection{Process to Label the Dataset}

We use the methodology given in the paper by Ren et al.~\cite{Making-Smart-Contract-Development-More-Secure-and-Easier} to find vulnerabilities in smart contracts. To understand the process of labeling the dataset, consider the example shown in Listing~\ref{lst:solidity-bug}. It shows a code snippet of a vulnerable contract from the SWC registry. Here, the constructor name should be \textbf{``Missing''}. However, accidentally, the smart contract developer writes it as \textbf{``IamMissing''}. Now, any node on the Ethereum blockchain can create a transaction that calls this function named \textbf{``IamMissing''} and can become the owner of this contract. This type of bug is known as \textbf{incorrect constructor name}.

\begin{lstlisting}[float,caption=Solidity Bug: Incorrect Constructor name at line~\ref{lst:buggy-line}, escapechar=\%, language=Solidity, label=lst:solidity-bug]
pragma solidity ^0.4.24;
contract Missing{
    address private owner;
    function IamMissing() public { %\label{lst:buggy-line}%
        owner = msg.sender;
    }
}
\end{lstlisting}

Due to the presence of false positives in the reports generated by the tools, we cannot depend on just one tool. We use majority voting, that is, at least 50\% of the tools must report the same vulnerability at the same location. For example, assume that tools T1, T2, and T3 can detect incorrect constructor name vulnerability. Suppose, at least two tools report that {incorrect constructor name} vulnerability is present at Line~\ref{lst:buggy-line}. Then we say that the contract is vulnerable for {incorrect constructor name}; otherwise, we say that the contract is not vulnerable. 

Algorithm~\ref{alg:label_dataset} shows the methodology to label the dataset. The following steps explain the methodology given in the algorithm~\ref{alg:label_dataset}:

\begin{enumerate}

\item First, collect the output of the selected tools for the smart contract to be analyzed (line~\ref{alg:collect_output_of_tool}).

\item Note the vulnerability name and line number for the vulnerability per tool (line~\ref{alg:note_line_vuln}). 

\item For the same vulnerability, if different tools show different locations, consider them as different warnings. We consider two warnings as the same only if the vulnerability name and location match for different tools (line~\ref{alg:increase_tool_count}).

\item The methodology says that the vulnerability exists at a location if more than 50\% tools confirm the same vulnerability at the exact location. In that case, label the location with the vulnerability (line~\ref{alg:start_threshold_loop} to~\ref{alg:end_threshold_loop}).

\end{enumerate}

\begin{algorithm}
\caption{Process to Label the Dataset}\label{alg:label_dataset}
\begin{algorithmic}[1]
\For{$tool \gets tool\_1$ to $tool\_n$}
\State Collect the output of $tool$ \label{alg:collect_output_of_tool}
\For{$vuln \gets vuln\_1$ to $vuln\_n$}
\State 1. Note the $line\_number$ for $vuln$ detected by $tool$ 
\label{alg:note_line_vuln}
\State 2. Increase the $tool\_count$ for $vuln$ for $line\_number$ detected
\label{alg:increase_tool_count}
\EndFor
\EndFor

\For{$vuln \gets vuln\_1$ to $vuln\_n$}
\label{alg:start_threshold_loop}
 \If{$tool\_count$ >= n/2 for $line\_number$}
        \State Vulnerability : $vuln$ is present at $line\_number$
    \EndIf

\EndFor
\label{alg:end_threshold_loop}
\end{algorithmic}
\end{algorithm}

\section{Results}

\begin{figure}[t]
  \centering
  \includegraphics[width=\linewidth]{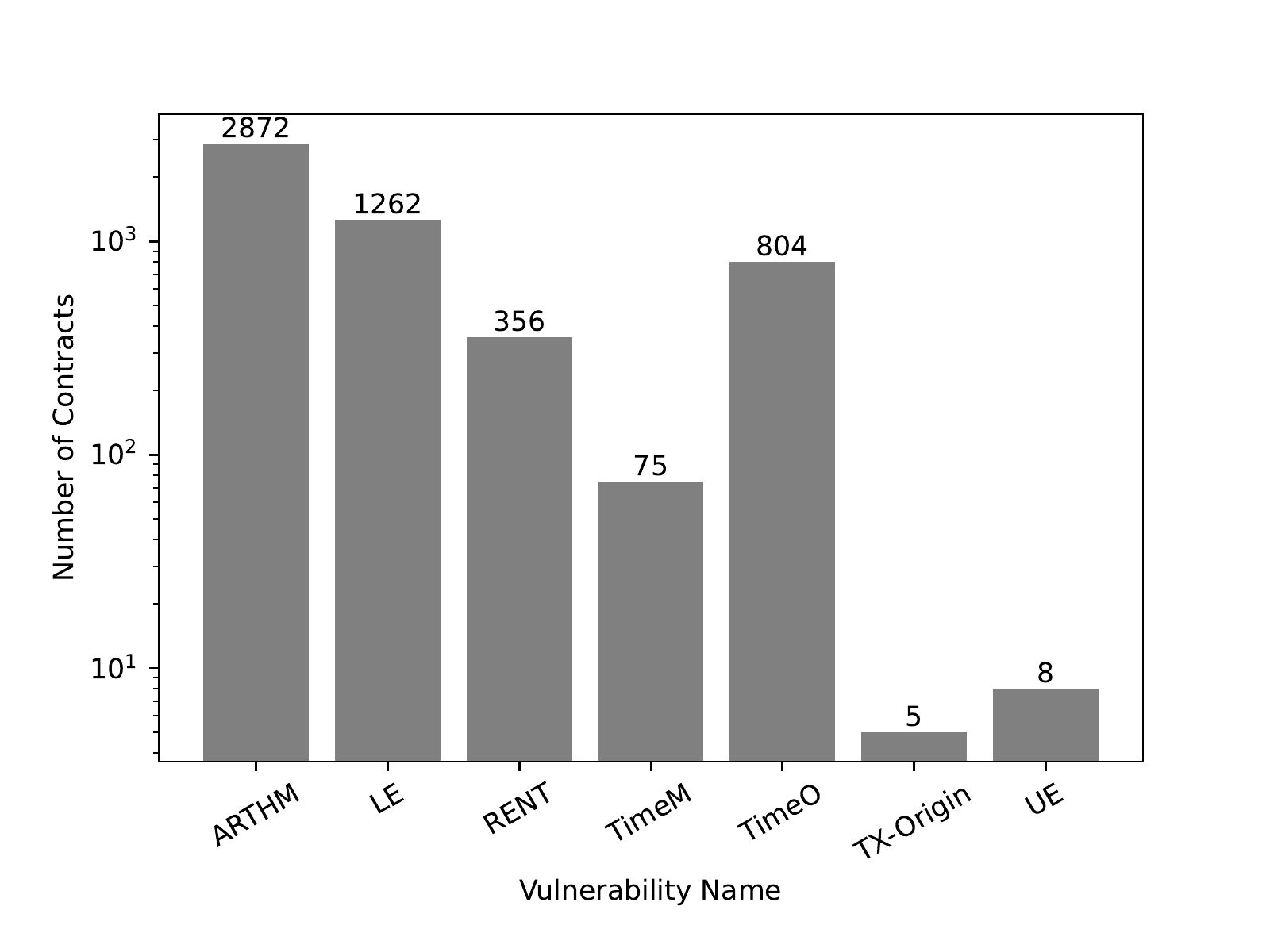}
  
  \caption{Number of unique Contracts having each Vulnerability (log-scale), Vulnerability Type DOS is not detected by majority voting}
  \label{fig:contracts_per_vuln}
\end{figure}

\begin{figure}[t]
  \centering
  \includegraphics[width=\linewidth]{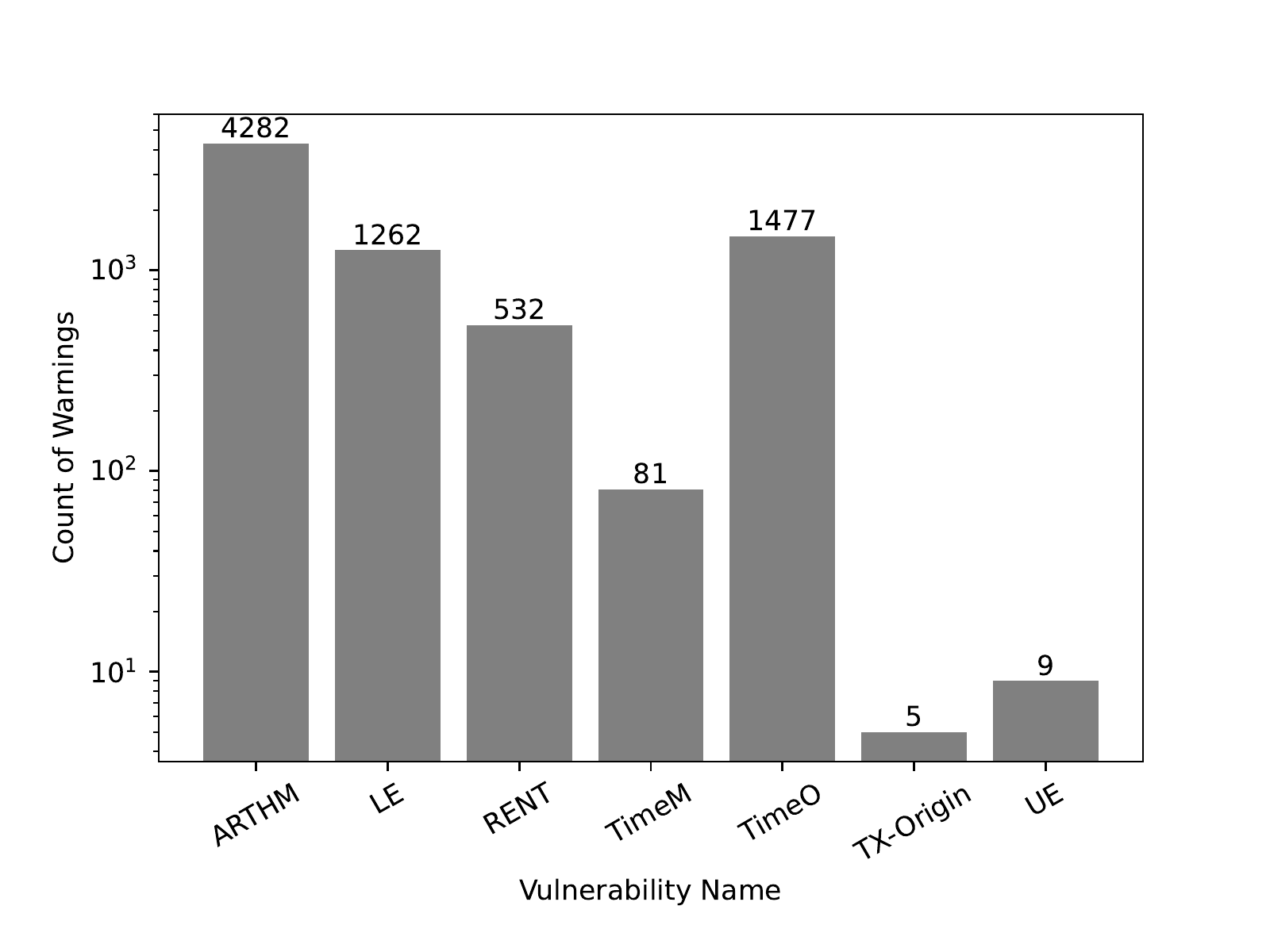}
  \caption{Count of Vulnerability Warnings according to the Majority of Tools (log-scale), Vulnerability Type DOS is not detected by majority voting}
  \label{fig:warnings_per_vuln}
\end{figure}

\begin{table}[t]
  \caption{Total Contracts with Unique Vulnerabilities}
  \label{table:total_contracts_with_uniq_vuln}
  \begin{tabular}{P{0.5\linewidth}P{0.3\linewidth}}
    \toprule
    \textbf{\#Unique Vulnerabilities} & \textbf{\#Contracts}  \\
    \midrule
    NONE & 2688 \\
    1 & 2912 \\
    2 & 1082 \\
    3 & 86 \\
    4 & 12 \\
\bottomrule

\end{tabular}

\end{table}

Table \ref{table:threshold} shows the number of tools that support each vulnerability. The threshold column in the table shows the majority rule. For example, consider Unhandled Exception (UE) vulnerability. Three tools support it. Its threshold is two. Hence, if two or more tools say that the contract is vulnerable for UE at the same location, we report it as a real vulnerability. We show the vulnerability only if greater than or equal to 50\% of the tools report that it is present. The idea of marking the vulnerability as real or not is taken from the paper by Ren et al. \cite{Making-Smart-Contract-Development-More-Secure-and-Easier}.

Figure~\ref{fig:contracts_per_vuln} shows the number of smart contracts with at least one occurrence of particular vulnerability according to the majority vote.
 For example, consider Reentrancy vulnerability. Table \ref{table:threshold} shows that Slither, Mythril, and Oyente support the detection reentrancy vulnerability. We say that the contract contains reentrancy only if two of these three tools report that it is present at the same location.

Figure~\ref{fig:warnings_per_vuln} shows the total warnings reported for each vulnerability according to the majority of tools. Please note that there can be multiple warnings in a single file for each vulnerability.

Table~\ref{table:total_contracts_with_uniq_vuln} shows contracts having exclusive number of vulnerabilities. According to our evaluation methodology, 2688 contracts have no vulnerabilities. 2912 contracts have only one type of vulnerability. Similarly, 1082 contracts have two  vulnerabilities each.

\section{Applications of The Dataset}
    \subsection{Evaluation of New Tools} 

    The Ethereum community can use our dataset for evaluating correctness and other parameters of newly designed tools. \mytool~comprises a diverse set of real-world annotated smart contracts. Hence, researchers can use \mytool~for evaluating new tools. 

    \subsection{Machine Learning based Tools}
    Vulnerability analysis tools based on Machine Learning needs dataset for training. \mytool~has \totalSCInOurDB labelled smart contracts. Hence, researchers can use \mytool~as a basis for learning or training of ML-based tools. 
    
    \subsection{Empirical Analysis of existing Tools}
    There are plenty of tools that analyze Ethereum smart contracts and detect different vulnerabilities. Smart contract developers face difficulty choosing the tool to analyze their smart contracts. There should be an empirical study that compares various tools and suggest which tools are better. Previous works \cite{Empirical-Evaluation-of-Smart-Contract-Testing:What-is-the-Best-Choice,Empirical-Review-of-Automated-Analysis-Tools-on-47587-Ethereum-Smart-Contracts} use dataset where the annotation is limited to a few thousand smart contracts. The effectiveness of these studies can be improved with \mytool.

\section{Challenges in creating the Annotated Dataset} \label{section:challenges}
Each tool depends on different packages, specific versions of these packages. Most of the time, different tools are not compatible with each other. That's why we create docker \cite{docker} containers for these tools and run our experiments.

Mythril tool is memory-intensive. It consumes more than 10 GB RAM, and hence we cannot run it in parallel. It is cumbersome to get results from Mythril.

\section{Threats to Validity} \label{section:threats-to-validity}
Each tool selected for evaluation is neither sound nor complete. There can be a lot of false positives or false negatives generated by a specific tool. However, we used the majority approach to say whether a vulnerability is present or not. But, this approach may fail if majority of the tools generate false positives or false negatives.

We can solve these issues by integrating more tools or manually reviewing detected vulnerabilities through crowdsourcing or other means. However, these approaches are time and resource-intensive, and we can do it incrementally.

\section{Conclusion and Future Work}

We collected 46K Ethereum smart contracts from Etherscan. However, due to challenges in section~\ref{section:challenges}, we could not label them all. We are working on labeling more contracts, and \mytool~will be updated regularly. 

We will add more tools to \mytool~and do a manual inspection of detected vulnerabilities in smart contracts through crowdsourcing or some other means in the future.


\bibliographystyle{ACM-Reference-Format}
\balance
\bibliography{reference}


\begin{thebibliography}{18}


\ifx \showCODEN    \undefined \def \showCODEN     #1{\unskip}     \fi
\ifx \showDOI      \undefined \def \showDOI       #1{#1}\fi
\ifx \showISBNx    \undefined \def \showISBNx     #1{\unskip}     \fi
\ifx \showISBNxiii \undefined \def \showISBNxiii  #1{\unskip}     \fi
\ifx \showISSN     \undefined \def \showISSN      #1{\unskip}     \fi
\ifx \showLCCN     \undefined \def \showLCCN      #1{\unskip}     \fi
\ifx \shownote     \undefined \def \shownote      #1{#1}          \fi
\ifx \showarticletitle \undefined \def \showarticletitle #1{#1}   \fi
\ifx \showURL      \undefined \def \showURL       {\relax}        \fi
\providecommand\bibfield[2]{#2}
\providecommand\bibinfo[2]{#2}
\providecommand\natexlab[1]{#1}
\providecommand\showeprint[2][]{arXiv:#2}

\bibitem[\protect\citeauthoryear{Atzei, Bartoletti, and Cimoli}{Atzei
  et~al\mbox{.}}{2017}]%
        {A-Survey-of-Attacks-on-Ethereum-Smart-Contracts}
\bibfield{author}{\bibinfo{person}{Nicola Atzei}, \bibinfo{person}{Massimo
  Bartoletti}, {and} \bibinfo{person}{Tiziana Cimoli}.}
  \bibinfo{year}{2017}\natexlab{}.
\newblock \showarticletitle{A Survey of Attacks on Ethereum Smart Contracts
  (SoK)}. In \bibinfo{booktitle}{\emph{Principles of Security and Trust}},
  \bibfield{editor}{\bibinfo{person}{Matteo Maffei} {and} \bibinfo{person}{Mark
  Ryan}} (Eds.). \bibinfo{publisher}{Springer Berlin Heidelberg},
  \bibinfo{address}{Berlin, Heidelberg}, \bibinfo{pages}{164--186}.
\newblock
\showISBNx{978-3-662-54455-6}


\bibitem[\protect\citeauthoryear{Chen, Pendleton, Njilla, and Xu}{Chen
  et~al\mbox{.}}{2020}]%
  {A-Survey-on-Ethereum-Systems-Security-Vulnerabilities-Attacks-and-Defenses}
\bibfield{author}{\bibinfo{person}{Huashan Chen}, \bibinfo{person}{Marcus
  Pendleton}, \bibinfo{person}{Laurent Njilla}, {and} \bibinfo{person}{Shouhuai
  Xu}.} \bibinfo{year}{2020}\natexlab{}.
\newblock \showarticletitle{A Survey on Ethereum Systems Security:
  Vulnerabilities, Attacks, and Defenses}.
\newblock \bibinfo{journal}{\emph{ACM Comput. Surv.}} \bibinfo{volume}{53},
  \bibinfo{number}{3}, Article \bibinfo{articleno}{67} (\bibinfo{date}{jun}
  \bibinfo{year}{2020}), \bibinfo{numpages}{43}~pages.
\newblock
\showISSN{0360-0300}
\urldef\tempurl%
\url{https://doi.org/10.1145/3391195}
\showDOI{\tempurl}


\bibitem[\protect\citeauthoryear{ConsenSys}{ConsenSys}{2022}]%
        {mythril}
\bibfield{author}{\bibinfo{person}{ConsenSys}.}
  \bibinfo{year}{2022}\natexlab{}.
\newblock \bibinfo{title}{Mythril}.
\newblock
\newblock
\urldef\tempurl%
\url{https://github.com/ConsenSys/mythril-classic}
\showURL{%
\tempurl}


\bibitem[\protect\citeauthoryear{Daian}{Daian}{2022}]%
        {dao}
\bibfield{author}{\bibinfo{person}{Phil Daian}.}
  \bibinfo{year}{2022}\natexlab{}.
\newblock \bibinfo{title}{Analysis of the dao exploit.}
\newblock
\newblock
\urldef\tempurl%
\url{http://hackingdistributed.com/ 2016/06/18/analysis-of-the-dao-exploit}
\showURL{%
\tempurl}


\bibitem[\protect\citeauthoryear{Docker}{Docker}{2013}]%
        {docker}
\bibfield{author}{\bibinfo{person}{Docker}.} \bibinfo{year}{Initial Release:
  2013}\natexlab{}.
\newblock \bibinfo{title}{Empowering App Development for Developers.}
\newblock
\newblock
\urldef\tempurl%
\url{https://www.docker.com}
\showURL{%
\tempurl}


\bibitem[\protect\citeauthoryear{Durieux, Ferreira, Abreu, and Cruz}{Durieux
  et~al\mbox{.}}{2020}]%
  {Empirical-Review-of-Automated-Analysis-Tools-on-47587-Ethereum-Smart-Contracts}
\bibfield{author}{\bibinfo{person}{Thomas Durieux},
  \bibinfo{person}{Jo\~{a}o~F. Ferreira}, \bibinfo{person}{Rui Abreu}, {and}
  \bibinfo{person}{Pedro Cruz}.} \bibinfo{year}{2020}\natexlab{}.
\newblock \showarticletitle{Empirical Review of Automated Analysis Tools on
  47,587 Ethereum Smart Contracts}. In \bibinfo{booktitle}{\emph{Proceedings of
  the ACM/IEEE 42nd International Conference on Software Engineering}} (Seoul,
  South Korea) \emph{(\bibinfo{series}{ICSE '20})}.
  \bibinfo{publisher}{Association for Computing Machinery},
  \bibinfo{address}{New York, NY, USA}, \bibinfo{pages}{530–541}.
\newblock
\showISBNx{9781450371216}
\urldef\tempurl%
\url{https://doi.org/10.1145/3377811.3380364}
\showDOI{\tempurl}


\bibitem[\protect\citeauthoryear{Etherscan}{Etherscan}{2022}]%
        {etherscan}
\bibfield{author}{\bibinfo{person}{Etherscan}.}
  \bibinfo{year}{2022}\natexlab{}.
\newblock \bibinfo{title}{The Ethereum Blockchain Explorer}.
\newblock
\newblock
\urldef\tempurl%
\url{https://etherscan.io}
\showURL{%
\tempurl}


\bibitem[\protect\citeauthoryear{Feist, Greico, and Groce}{Feist
  et~al\mbox{.}}{2019}]%
        {slither}
\bibfield{author}{\bibinfo{person}{Josselin Feist}, \bibinfo{person}{Gustavo
  Greico}, {and} \bibinfo{person}{Alex Groce}.}
  \bibinfo{year}{2019}\natexlab{}.
\newblock \showarticletitle{Slither: A Static Analysis Framework for Smart
  Contracts}. In \bibinfo{booktitle}{\emph{Proceedings of the 2nd International
  Workshop on Emerging Trends in Software Engineering for Blockchain}}
  (Montreal, Quebec, Canada) \emph{(\bibinfo{series}{WETSEB '19})}.
  \bibinfo{publisher}{IEEE Press}, \bibinfo{pages}{8–15}.
\newblock
\urldef\tempurl%
\url{https://doi.org/10.1109/WETSEB.2019.00008}
\showDOI{\tempurl}


\bibitem[\protect\citeauthoryear{Ferreira, Cruz, Durieux, and Abreu}{Ferreira
  et~al\mbox{.}}{2020}]%
        {smartbugs}
\bibfield{author}{\bibinfo{person}{Jo\~{a}o~F. Ferreira},
  \bibinfo{person}{Pedro Cruz}, \bibinfo{person}{Thomas Durieux}, {and}
  \bibinfo{person}{Rui Abreu}.} \bibinfo{year}{2020}\natexlab{}.
\newblock \showarticletitle{SmartBugs: A Framework to Analyze Solidity Smart
  Contracts}. In \bibinfo{booktitle}{\emph{Proceedings of the 35th IEEE/ACM
  International Conference on Automated Software Engineering}} (Virtual Event,
  Australia) \emph{(\bibinfo{series}{ASE '20})}.
  \bibinfo{publisher}{Association for Computing Machinery},
  \bibinfo{address}{New York, NY, USA}, \bibinfo{pages}{1349–1352}.
\newblock
\showISBNx{9781450367684}
\urldef\tempurl%
\url{https://doi.org/10.1145/3324884.3415298}
\showDOI{\tempurl}


\bibitem[\protect\citeauthoryear{Luu, Chu, Olickel, Saxena, and Hobor}{Luu
  et~al\mbox{.}}{2016}]%
        {oyente}
\bibfield{author}{\bibinfo{person}{Loi Luu}, \bibinfo{person}{Duc-Hiep Chu},
  \bibinfo{person}{Hrishi Olickel}, \bibinfo{person}{Prateek Saxena}, {and}
  \bibinfo{person}{Aquinas Hobor}.} \bibinfo{year}{2016}\natexlab{}.
\newblock \showarticletitle{Making Smart Contracts Smarter}. In
  \bibinfo{booktitle}{\emph{Proceedings of the 2016 ACM SIGSAC Conference on
  Computer and Communications Security}} (Vienna, Austria)
  \emph{(\bibinfo{series}{CCS '16})}. \bibinfo{publisher}{Association for
  Computing Machinery}, \bibinfo{address}{New York, NY, USA},
  \bibinfo{pages}{254–269}.
\newblock
\showISBNx{9781450341394}
\urldef\tempurl%
\url{https://doi.org/10.1145/2976749.2978309}
\showDOI{\tempurl}


\bibitem[\protect\citeauthoryear{Miller, Cai, and Jha}{Miller
  et~al\mbox{.}}{2018}]%
        {Smart-Contracts-and-Opportunities-for-Formal-Methods}
\bibfield{author}{\bibinfo{person}{Andrew Miller}, \bibinfo{person}{Zhicheng
  Cai}, {and} \bibinfo{person}{Somesh Jha}.} \bibinfo{year}{2018}\natexlab{}.
\newblock \showarticletitle{Smart Contracts and Opportunities for Formal
  Methods}. In \bibinfo{booktitle}{\emph{Leveraging Applications of Formal
  Methods, Verification and Validation. Industrial Practice}},
  \bibfield{editor}{\bibinfo{person}{Tiziana Margaria} {and}
  \bibinfo{person}{Bernhard Steffen}} (Eds.). \bibinfo{publisher}{Springer
  International Publishing}, \bibinfo{address}{Cham},
  \bibinfo{pages}{280--299}.
\newblock
\showISBNx{978-3-030-03427-6}


\bibitem[\protect\citeauthoryear{Nikoli\'{c}, Kolluri, Sergey, Saxena, and
  Hobor}{Nikoli\'{c} et~al\mbox{.}}{2018}]%
        {maian}
\bibfield{author}{\bibinfo{person}{Ivica Nikoli\'{c}}, \bibinfo{person}{Aashish
  Kolluri}, \bibinfo{person}{Ilya Sergey}, \bibinfo{person}{Prateek Saxena},
  {and} \bibinfo{person}{Aquinas Hobor}.} \bibinfo{year}{2018}\natexlab{}.
\newblock \showarticletitle{Finding The Greedy, Prodigal, and Suicidal
  Contracts at Scale}. In \bibinfo{booktitle}{\emph{Proceedings of the 34th
  Annual Computer Security Applications Conference}} (San Juan, PR, USA)
  \emph{(\bibinfo{series}{ACSAC '18})}. \bibinfo{publisher}{Association for
  Computing Machinery}, \bibinfo{address}{New York, NY, USA},
  \bibinfo{pages}{653–663}.
\newblock
\showISBNx{9781450365697}
\urldef\tempurl%
\url{https://doi.org/10.1145/3274694.3274743}
\showDOI{\tempurl}


\bibitem[\protect\citeauthoryear{Ren, Ma, Yin, Fu, Li, Chang, and Jiang}{Ren
  et~al\mbox{.}}{2021a}]%
        {Making-Smart-Contract-Development-More-Secure-and-Easier}
\bibfield{author}{\bibinfo{person}{Meng Ren}, \bibinfo{person}{Fuchen Ma},
  \bibinfo{person}{Zijing Yin}, \bibinfo{person}{Ying Fu},
  \bibinfo{person}{Huizhong Li}, \bibinfo{person}{Wanli Chang}, {and}
  \bibinfo{person}{Yu Jiang}.} \bibinfo{year}{2021}\natexlab{a}.
\newblock \showarticletitle{Making Smart Contract Development More Secure and
  Easier}. In \bibinfo{booktitle}{\emph{Proceedings of the 29th ACM Joint
  Meeting on European Software Engineering Conference and Symposium on the
  Foundations of Software Engineering}} (Athens, Greece)
  \emph{(\bibinfo{series}{ESEC/FSE 2021})}. \bibinfo{publisher}{Association for
  Computing Machinery}, \bibinfo{address}{New York, NY, USA},
  \bibinfo{pages}{1360–1370}.
\newblock
\showISBNx{9781450385626}
\urldef\tempurl%
\url{https://doi.org/10.1145/3468264.3473929}
\showDOI{\tempurl}


\bibitem[\protect\citeauthoryear{Ren, Yin, Ma, Xu, Jiang, Sun, Li, and Cai}{Ren
  et~al\mbox{.}}{2021b}]%
  {Empirical-Evaluation-of-Smart-Contract-Testing:What-is-the-Best-Choice}
\bibfield{author}{\bibinfo{person}{Meng Ren}, \bibinfo{person}{Zijing Yin},
  \bibinfo{person}{Fuchen Ma}, \bibinfo{person}{Zhenyang Xu},
  \bibinfo{person}{Yu Jiang}, \bibinfo{person}{Chengnian Sun},
  \bibinfo{person}{Huizhong Li}, {and} \bibinfo{person}{Yan Cai}.}
  \bibinfo{year}{2021}\natexlab{b}.
\newblock \showarticletitle{Empirical Evaluation of Smart Contract Testing:
  What is the Best Choice?}. In \bibinfo{booktitle}{\emph{Proceedings of the
  30th ACM SIGSOFT International Symposium on Software Testing and Analysis}}
  (Virtual, Denmark) \emph{(\bibinfo{series}{ISSTA 2021})}.
  \bibinfo{publisher}{Association for Computing Machinery},
  \bibinfo{address}{New York, NY, USA}, \bibinfo{pages}{566–579}.
\newblock
\showISBNx{9781450384599}
\urldef\tempurl%
\url{https://doi.org/10.1145/3460319.3464837}
\showDOI{\tempurl}


\bibitem[\protect\citeauthoryear{SpankChain}{SpankChain}{2022}]%
        {spankchain}
\bibfield{author}{\bibinfo{person}{SpankChain}.}
  \bibinfo{year}{2022}\natexlab{}.
\newblock \bibinfo{title}{We got spanked: What we know so far.}
\newblock
\newblock
\urldef\tempurl%
\url{https://medium.
  com/spankchain/we-got-spanked-what-we-know-so-far-d5ed3a0f38fe}
\showURL{%
\tempurl}


\bibitem[\protect\citeauthoryear{Tikhomirov, Voskresenskaya, Ivanitskiy,
  Takhaviev, Marchenko, and Alexandrov}{Tikhomirov et~al\mbox{.}}{2018}]%
        {smartcheck}
\bibfield{author}{\bibinfo{person}{Sergei Tikhomirov},
  \bibinfo{person}{Ekaterina Voskresenskaya}, \bibinfo{person}{Ivan
  Ivanitskiy}, \bibinfo{person}{Ramil Takhaviev}, \bibinfo{person}{Evgeny
  Marchenko}, {and} \bibinfo{person}{Yaroslav Alexandrov}.}
  \bibinfo{year}{2018}\natexlab{}.
\newblock \showarticletitle{SmartCheck: Static Analysis of Ethereum Smart
  Contracts}. In \bibinfo{booktitle}{\emph{Proceedings of the 1st International
  Workshop on Emerging Trends in Software Engineering for Blockchain}}
  (Gothenburg, Sweden) \emph{(\bibinfo{series}{WETSEB '18})}.
  \bibinfo{publisher}{Association for Computing Machinery},
  \bibinfo{address}{New York, NY, USA}, \bibinfo{pages}{9–16}.
\newblock
\showISBNx{9781450357265}
\urldef\tempurl%
\url{https://doi.org/10.1145/3194113.3194115}
\showDOI{\tempurl}


\bibitem[\protect\citeauthoryear{Torres, Sch\"{u}tte, and State}{Torres
  et~al\mbox{.}}{2018}]%
        {osiris}
\bibfield{author}{\bibinfo{person}{Christof~Ferreira Torres},
  \bibinfo{person}{Julian Sch\"{u}tte}, {and} \bibinfo{person}{Radu State}.}
  \bibinfo{year}{2018}\natexlab{}.
\newblock \showarticletitle{Osiris: Hunting for Integer Bugs in Ethereum Smart
  Contracts}. In \bibinfo{booktitle}{\emph{Proceedings of the 34th Annual
  Computer Security Applications Conference}} (San Juan, PR, USA)
  \emph{(\bibinfo{series}{ACSAC '18})}. \bibinfo{publisher}{Association for
  Computing Machinery}, \bibinfo{address}{New York, NY, USA},
  \bibinfo{pages}{664–676}.
\newblock
\showISBNx{9781450365697}
\urldef\tempurl%
\url{https://doi.org/10.1145/3274694.3274737}
\showDOI{\tempurl}


\bibitem[\protect\citeauthoryear{Tsankov, Dan, Drachsler-Cohen, Gervais,
  B\"{u}nzli, and Vechev}{Tsankov et~al\mbox{.}}{2018}]%
        {securify}
\bibfield{author}{\bibinfo{person}{Petar Tsankov}, \bibinfo{person}{Andrei
  Dan}, \bibinfo{person}{Dana Drachsler-Cohen}, \bibinfo{person}{Arthur
  Gervais}, \bibinfo{person}{Florian B\"{u}nzli}, {and} \bibinfo{person}{Martin
  Vechev}.} \bibinfo{year}{2018}\natexlab{}.
\newblock \showarticletitle{Securify: Practical Security Analysis of Smart
  Contracts}. In \bibinfo{booktitle}{\emph{Proceedings of the 2018 ACM SIGSAC
  Conference on Computer and Communications Security}} (Toronto, Canada)
  \emph{(\bibinfo{series}{CCS '18})}. \bibinfo{publisher}{Association for
  Computing Machinery}, \bibinfo{address}{New York, NY, USA},
  \bibinfo{pages}{67–82}.
\newblock
\showISBNx{9781450356930}
\urldef\tempurl%
\url{https://doi.org/10.1145/3243734.3243780}
\showDOI{\tempurl}


\end{thebibliography}

\end{document}